\documentclass[proceedings, preprint]{rmaa}

\usepackage[utf8]{inputenc}   
\usepackage[english]{babel}   
\usepackage{graphicx}        
\usepackage[free-standing-units]{siunitx}
\usepackage{paralist}        
\usepackage{xspace}          
\usepackage[usenames,dvipsnames]{xcolor} 
\usepackage[pdftex]{hyperref}
\hypersetup{colorlinks,urlcolor=MidnightBlue,linkcolor=MidnightBlue,citecolor=MidnightBlue}

\SetYear{2020}
\SetConfTitle{Sixth Workshop on robotic autonomous observatories}

\title{Technical and software upgrades completed and planned at OARPAF} 

\author{
  D.~Ricci\altaffilmark{1},
  L.~Cabona\altaffilmark{2,3},
  A.~La~Camera\altaffilmark{4},
  C.~Righi\altaffilmark{2,5},
  F.~Nicolosi\altaffilmark{6} and
  S.~Tosi\altaffilmark{5,6}
  }

\altaffiltext{1}{
  INAF -- Osservatorio Astronomico di Padova,
  Vicolo dell'Osservatorio 5, 35122 Padua, Italy.
}

\altaffiltext{2}{
  INAF -- Osservatorio Astronomico di Brera,
  via Bianchi 46, 23807 Merate, Italy.
}

\altaffiltext{3}{
  Universit\`a degli Studi dell'Insubria,
  via Valleggio 11, 22100 Como, Italy.
}

\altaffiltext{4}{
  Teiga srls, 
  Viale Brigate Partigiane 16, 16129 Genoa, Italy.
}

\altaffiltext{5}{
  Istituto Nazionale di Fisica Nucleare -- Sezione di Genova, 
  Via Dodecaneso 33, 16146 Genoa, Italy. 
}

\altaffiltext{6}{
  Dipartimento di Fisica -- Universit\`a degli studi di Genova,
  Via Dodecaneso 33, 16146 Genoa, Italy.
}

\shortauthor{Ricci et al.}
\shorttitle{Technical and software upgrades at OARPAF}
\listofauthors{ D.~Ricci, L.~Cabona, C.~Righi, A.~La~Camera, F.~Nicolosi, S.~Tosi}
\indexauthor{Ricci, D.}
\indexauthor{Cabona, L.}
\indexauthor{Righi, C.}
\indexauthor{La~Camera, A.}
\indexauthor{Nicolosi, F.}
\indexauthor{Tosi S.}

\abstract{ We present technical, instrumental, and software upgrades
  completed and planned at astronomical observatory called 
  ``Osservatorio Astronomico Regionale Parco Antola, Fascia'' (OARPAF), 
  hosting an 80\centi\meter, alt-az Cassegrain-Nasmyth telescope. 
  The observatory, located in the Ligurian Apennines, can currently be operated
  either for scientific (photometry camera) or amateur (ocular)
  observations, by switching the tertiary mirror between the two
  Nasmyth foci using a manual handle.  The main scientific
  observational topics are related up to now to exoplanetary transits,
  QSOs, and gravitationally lensed quasars, and results are being
  recently published.  A remotization and robotization strategy of the
  entire structure (telescope, dome, instruments, sensors and
  monitoring) have been set up and it is in progress.  We report the
  current upgrades, mainly related for what concerns the ``hardware''
  side to the robotization of the dome.  On the instrumentation side,
  a new modular support for instruments with spectrophotometric
  capabilities is on a preliminary design phase, improving the
  telescope performances and broadening the potential science
  fields. In this framework, the procurement of spectrophotometric
  material has started.  On the software side, an innovative web-based
  software relying on websockets and \texttt{node.js} can already be
  used to control the camera, and it will be extended to manage the
  other components of the instrument, of the observatory, and of the
  image database storage. }

\resumen{Presentamos actualizaciones t\'ecnicas, instrumentales y de software,
desarrolladas y planeadas en el observatorio astronómico denominado 
Osservatorio Astronomico Regionale Parco Antola, Fascia (OARPAF), que 
alberga un telescopio alt-azimutal Cassegrain-Nasmyth de 80 cm. El 
observatorio, ubicado en los Apeninos de Liguria, actualmente puede ser 
operado ya sea para uso cient\'ifico por profesionales (con c\'amara 
fotom\'etrica) o para observaciones por aficionados (usando oculares), 
simplemente cambiando el espejo terciario entre los dos focos Nasmyth de 
manera manual. Los principales campos de observaci\'on cient\'ifica 
est\'an relacionados con tr\'ansitos de exoplanetas, cu\'asares y lentes 
gravitacionales relacionadas con cu\'asares, habi\'endose publicado 
resultados recientemente. Una estrategia de remotizaci\'on y 
robotizaci\'on del observatorio ha permitido configurar toda la 
estructura (telescopio, c\'upula, instrumentos, sensores y aparatos de 
monitorizaci\'on) implement\'andose constantes mejoras. Presentamos en 
este trabajo las actualizaciones actuales, principalmente relacionadas 
con lo que concierne al lado hardware de la robotizaci\'on de la 
c\'upula por el lado de la instrumentaci\'on, as\'i como un nuevo 
soporte modular para instrumentci\'on con capacidades 
espectrofotom\'etrícas, que se encuentra en una fase de dise{\~n}o 
preliminar, pudiendo mejorar el rendimiento del telescopio y ampliando 
los potenciales campos cient\'ificos de investigaci\'on. En este marco, 
ha comenzado la adquisici\'on de material espectrofotom\'etrico. Por 
otro lado, en cuanto al software, se ha desarrollado un innovador 
software basado en web que, usando websockets y node.js ya se puede 
utilizar para controlar la c\'amara, y que se ampliar\'a para poder 
administrar los otros instrumentos del observatorio y para el 
almacenamiento de la base de datos de im\'agenes.
}

\addkeyword{Observational astronomy:Astronomical techniques:Spectroscopy}
\addkeyword{Observational astronomy:Astronomical techniques:Photometry}
\addkeyword{Observational astronomy:Astronomical instrumentation}
\addkeyword{Observational astronomy:Astronomical methods:Computational astronomy:Astronomy software}
\addkeyword{Observational astronomy:Astronomical methods:Computational astronomy:Computational methods:Astronomy database}

\newcommand{\stl}{\textsc{sbig stl 11000m} camera\xspace}
\newcommand{\stx}{\textsc{sbig stx 16801} camera\xspace}
\newcommand{\atik}{\textsc{atik 11000}\xspace}
\newcommand{\lhires}{\textsc{lhires iii}\xspace}
\newcommand{\flechas}{\textsc{flechas}\xspace}
\newcommand{\node}{\texttt{node.js}\xspace}

\begin{document}
\maketitle

\section{Overview}
\label{sec:overview}

\begin{figure}[t]
\centering
  \includegraphics[width=\columnwidth]{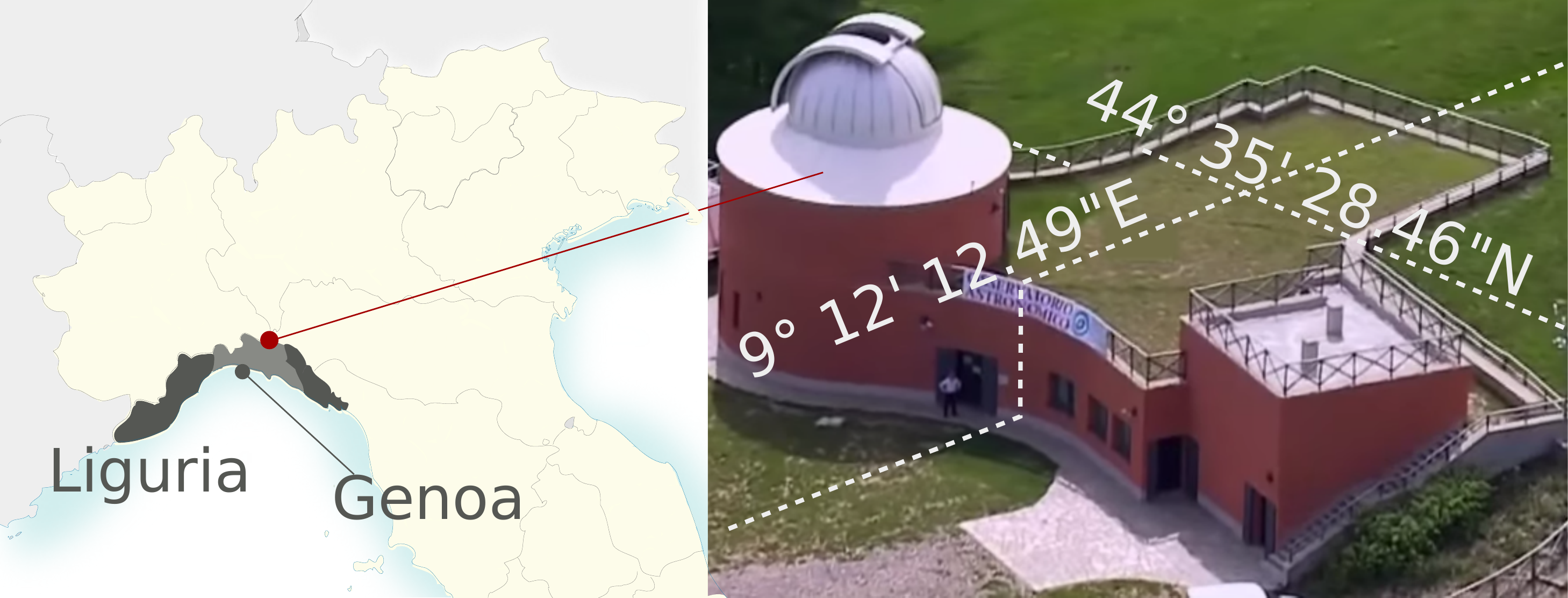}
  \caption{Location of the 80\centi\meter\ OARPAF telescope, at 
  1450\meter~a.s.l. within the Antola Regional Reserve. Coordinates 
  are: 44\degree~35'~28.46''~N, 9\degree~12'~12.49"E.}
  \label{fig:oarpaf}
\end{figure}

The astronomical observatory called ``Osservatorio Astronomico Regionale Parco 
Antola, Fascia'' (OARPAF), is realized in the Northern Italy, near Mt. Antola 
(Fig.~\ref{fig:oarpaf}), at 1450\meter~a.s.l. within the Antola Regional Reserve. 
The telescope is a Cassegrain-Nasmyth, with 80\centi\meter\ of primary diameter, 
alt-az mounted, and designed by the Astelco company to foresee a double Nasmyth 
focal stations. 
The first one, provided with a field derotator, is dedicated to scientific 
observations; the second one is dedicated to ocular observations by amateurs
\citep{2012ASInC...7....7F}.  The scientific Nasmyth focus is currently equipped 
with an air-cooled \stl, provided with an internal filter wheel with 
standard Johnson-Cousins $UBVRI$ filters.
Currently, the telescope is used for imaging/photometry by using this camera 
for several follow-up projects \citep{2016NCimC..39..284R, 2017PASP..129f4401R}.

The observatory will be remotized in the next future.  To reach this
goal, a planned stop of the scientific activities is allowing to
proceed with dome upgrades, telescope maintenance, and improvements of
Local Area Network (LAN) and internet connection.

We also decided to substitute the current CCD device with a
multi-functional instrument with superior capabilities, providing
imaging/photometry, long slit spectroscopy, échelle spectroscopy and a
``service'' focal station for further implementation.  Material to
reach this goal is being procured, and the \stl will be reused
for the project.

The instrument control software based on modern web technologies in
order to ease remote control, scalability, and maintenance.

In this proceedings we discuss the upgrades of the observatory
(Sec.~\ref{sec:oarpaf}), we describe the preliminary design of the new
instrument (Sec.~\ref{sec:cerbero}), and finally we give some detail of
the instrument control software (Sec.~\ref{sec:software}).
Conclusions are given in (Sec.~\ref{sec:conclusions}).

\section{Observatory status and upgrades}
\label{sec:oarpaf}

In recent years, the observatory suffered from several issues that
limited its functionalities.

The dome was initially not foreseen for remotization.  A custom
solution based on Raspberry~Pi~+~Arduino, with a command line Python
script, is used as interface between the Telescope Control
Software (TCS) and the dome to lock the rotation.  However, this
system do not control the eyelid, and recent bad weather conditions
stressed the structure so that it also accounts with water
infiltration.
For all these reasons, a new waterproof dome linked to the TCS
software, and with a doubly-motorized eyelid security system is
being installed by the Gambato company and financed by the University
of Genoa.

The Astelco company was also contacted in order to proceed with major
maintenance activities: cleaning of the primary mirror, optical
alignment, tests of the hydraulic pistons of the primary telescope
petals.  The proprietary TCS, \texttt{AstelOS}, will be also updated
to the latest version. This update will be financed by the Antola
Regional Reserve.

Malfunctions due to a thunderbolt affected the LAN switches, as well as
unstable internet connection, and slowed down the remotization efforts. 
Moreover, the \stl is power-connected to an electric socket placed in the
fork of the telescope, and USB-connected to a commercial pc inside the
fork.  This pc is accessed via Ethernet directly from a commercial
router through a home network.
In order to upgrade the network connections, the local administration 
allocated European and Regional funds to provide
optical fiber and high-speed internet access to the municipalities and
the public structures of the area surrounding the observatory.  
Moreover, the Antola Regional Reserve already procured the material 
for full implementation of a Gigabit LAN in the observatory. 
This update will be realized by the Liguria Digitale company.

\section{Instrumentation status and upgrades}
\label{sec:cerbero}

\begin{figure}[t]
  \includegraphics[width=\columnwidth]{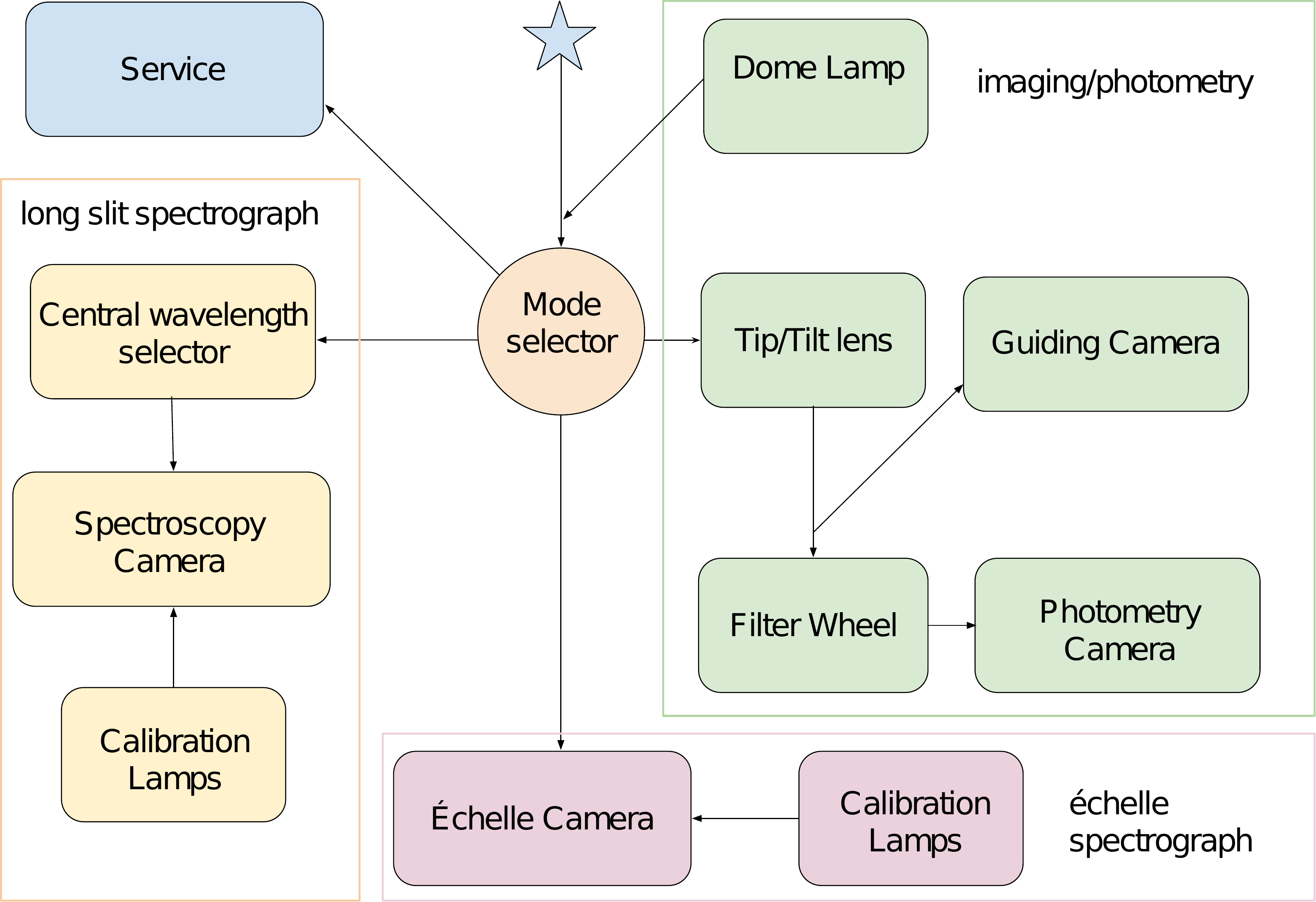}
  \caption{Preliminary design of the new OARPAF instrument. The mode 
  selector reflects the incoming light to one of the 3+1 instruments: 
  the imaging/photometry devices (in green, at the right), the échelle 
  spectrograph (in pink, at the bottom), the long slit spectrograph 
  (in yellow, at the left), and the ``service'' future instrument 
  (in blue, in the top-left). A further implementation foresees 
  a dome lamp for calibration. }
  \label{fig:cerbero}
\end{figure}

Currently, scientific activities at OARPAF are carried out using the
\stl operating at the derotated Nasmyth focus.
The observatory upgrades are the opportunity to improve scientific
activities at OARPAF.  To reach this goal, we plan to realize a new
spectrophotometric instrument at the derotated Nasmyth focus.  The
preliminary design foresees a flange interface, in order to provide
3+1 focal stations via a linear motor mode selector provided with
flat, 45\degree\ mirrors:
\begin{inparaenum}
\item imaging/photometry mode;
\item long slit spectroscopy mode;
\item échelle spectroscopy mode;
\item ``service'' mode.
\end{inparaenum}
All configurations use the derotated, on axis, embedded guiding camera
in the imaging/photometry path.

\subsection{Imaging/photometry mode}
\label{sec:imag-mode}

This is the main use mode of the new instrument.  The light proceeds
on a straight path from the telescope until a \textsc{sbig ao-x}
Tip/Tilt lens able to act up to a 10\hertz\ rate.  After this lens,
the embedded, on axis \textsc{sbig stx}-guider camera is placed just
before the 50\milli\meter\ \textsc{sbig fw-7} filter wheel provided
with $UBVRI$+$H\alpha$ filters and a free position.  This
$648 \times 486$, 7.4\micro\meter\ pixel guiding camera, placed in a
way that avoids vignetting, is provided with a $0.7\times$ focal
reducer, and is used to send offset information both to the telescope
and to the Tip/Tilt lens, if on.  Finally, the light focuses on the
$4096 \times 4096$, 9\micro\meter\ pixels array \stx with a Class 1
CCD for imaging and photometry on a 20\arcmin\ Field of View.

\subsection{Long slit spectroscopy mode}
\label{sec:long-slit-spectr}

This mode requires a flat mirror of the mode selector to address the
light on the slit of the \lhires spectrograph.  A part of the light
proceeds straight to the guiding camera on the imager and it is not
deviated by the flat mirror.  The remaining light passes through the
\lhires slit and the 1200 lines/\milli\meter\ diffraction grating,
focusing a spectral range of $\approx$140\nano\meter\ on the 4002
pixels of the \stl, that will be reused.  The central
wavelength can be selected by using a micrometer screw that tilts the
grating. This screw will be motorized in order to provide remote
control.  We foresee a resolution of $R\sim 5800$ in the
visible band (450--750\nano\meter).  A standard \textsc{spox} module
allows remote control of the embedded Neon-Argon calibration lamps.

\begin{figure}[t]
\centering
  \includegraphics[width=\columnwidth]{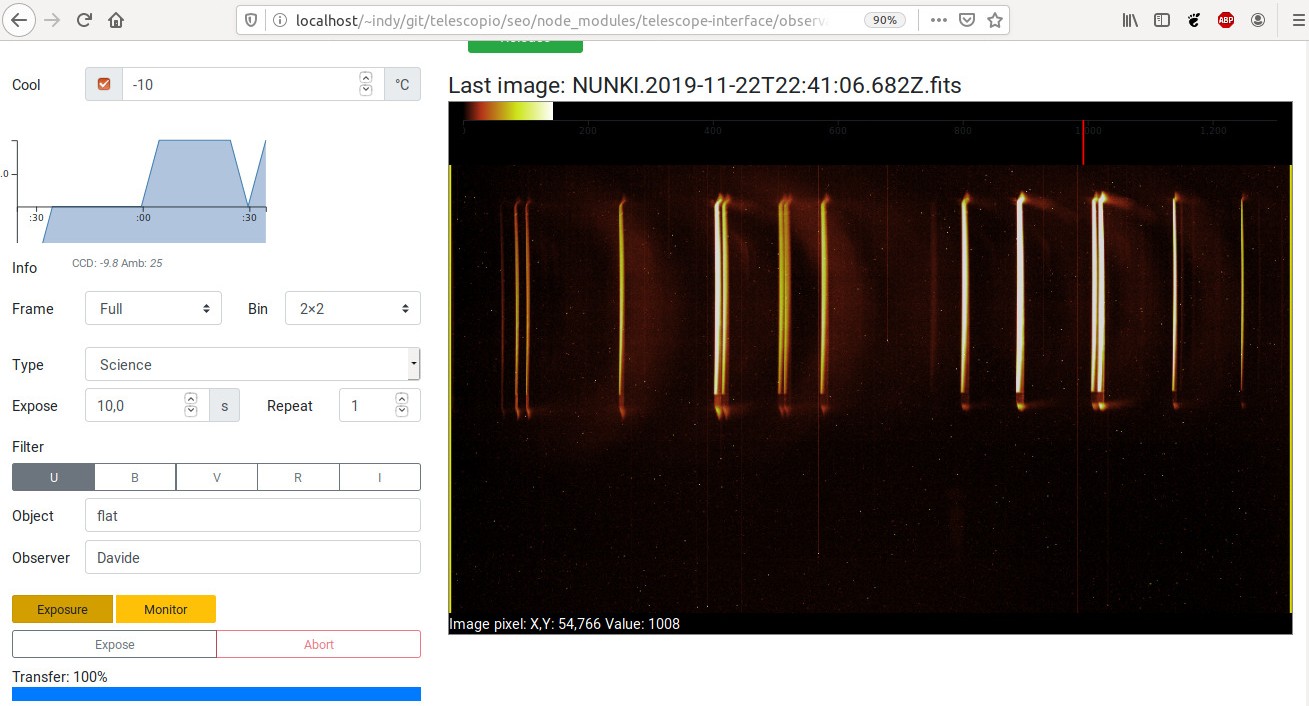}
  \caption{Preliminary design of the web interface for the control of 
  SBIG cameras. In this image, the \stl is used to take a 
  Neon-Argon lamp calibration frame for the \lhires.  }
  \label{fig:interface}
\end{figure}

\subsection{Échelle spectroscopy mode}
\label{sec:echelle-spectr-mode}

This mode requires a flat mirror to address the light on the head of a
\flechas échelle spectrograph, containing a Th-Ar calibration lamp and
supporting a 15\meter\ long optical fiber.  As in the case of the
long slit spectroscopy mode, part of the light is foreseen to proceed
straight to the guiding camera of the imager, while the light deviated
by the flat mirror is focused on a \atik camera on the \flechas at a
$R\sim 9300$ resolution.

\subsection{``Service'' mode}
\label{sec:service-mode}

This mode requires a flat mirror to address the light on focal station
for generic purposes.  Also in this case, part of the light is
required to proceed straight to the guiding camera on the imager. The
part of the light deviated by the flat mirror is focused on the
service focal station, for further instrument implementations,
maintenance operations, or, for example, astrophotography with a
digital reflex camera.

\section{Software status and upgrades}
\label{sec:software}

\begin{figure*}[t]
\centering
  \includegraphics[width=0.9\textwidth]{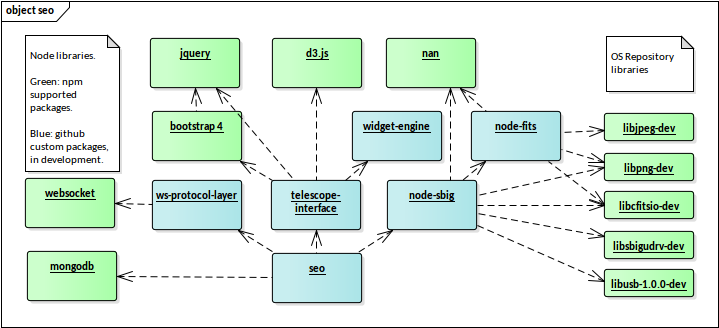}
  \caption{Class diagram of the \node test software. In blue, the custom 
  packages under development and available via GitHub (see text). In green, 
  standard packages available via the \texttt{npm} package manager.}
  \label{fig:software}
\end{figure*}

OARPAF telescope is currently operated via \texttt{ssh} connection and
web interface (see Fig.~\ref{fig:interface}), by using a laptop 
ethernet-connected to a local modem/router/switch, which provides 
both internet connection, and LAN access to several devices:
\begin{inparaenum}
\item the dome Raspberry Pi via \texttt{ssh} to move the dome;
\item the TCS pc via \texttt{ssh} to move the telescope;
\item the PC in the telescope fork via web browser and websocket to
  control the \stl detector and filter wheel.
\end{inparaenum}

In particular, we implemented the detector control software using
\node, a server-side \texttt{javascript} runtime environment.  \node
allows to extend the \texttt{javascript} engine built-in functions and
objects with new custom function and objects written in \texttt{C++}
and dynamically linked to its interpreter.

In order to test the capabilities of this technology for our purposes,
the \stl \node driver and FITS file manager were developed in two
separate modules by using this paradigm, while the web interface and
visualization was implemented thanks to standard and widely used
\texttt{html5} and \texttt{javascript} libraries, plus \texttt{WebGL}
(Fig.~\ref{fig:software}). All the developed packages are available 
in our GitHub repository \citep{github}.

In the framework of OARPAF observatory and instrumentation upgrades,
we want to use \node and web technologies to develop the control
software of the new instrument and the related detectors.  Dome and TCS
will also be directly interfaced with the instrument software by using
specific modules. Finally, the \texttt{javascript}-based, non-relational
\texttt{mongodb} will be used as database management system to archive
instrument status, logs, and FITS file position in a RAID-5 storage
device \citep{2014RMxAC..45...75S}.

This allows to control instrument, detector, telescope, and dome, with
a simple web browser, simplifying the remotization process, taking
advantage from the widely support of the components by professional
companies and the user community at large, and from the fact that web
browser are indeed the most reliable end-user component of every
OS-based device.

In order to ensure the security of web sessions and transmitted data, 
we plan to use a virtual private network (VPN), thanks to the 
implementation performed by the Liguria Digitale company.

\section{Conclusions}
\label{sec:conclusions}

In this proceedings we presented the current observatory,
instrumentation, and software status at OARPAF, and upgrades are in
progress in order to overcome current limitations.
The dome is being substituted, the telescope maintenance is already
programmed in next months, as well as the new LAN and internet
connection set up.
A new instrument providing photometry, long slit, and échelle
spectroscopy is in a preliminary design phase.
A web-based control software to control this new instrument, its
detectors, the telescope, and the dome, has been recently tested on a
software prototype.
%

\section*{Acknowledgements}
\label{sec:acknowledgements}

We thank Antola Regional Reserve for financial support.

\bibliography{biblio}{}

\begin{thebibliography}
\expandafter\ifx\csname natexlab\endcsname\relax\def\natexlab#1{#1}\fi
\expandafter\ifx\csname href\endcsname\relax
  \def\href#1#2{}\fi
\expandafter\ifx\csname urllinklabel\endcsname\relax
  \def\urllinklabel{[LINK]}\fi
\expandafter\ifx\csname adsurllinklabel\endcsname\relax
  \def\adsurllinklabel{[ADS]}\fi

\bibitem[{{Federici} {et~al.}(2012){Federici}, {Arduino}, {Riva}, \&
  {Zerbi}}]{2012ASInC...7....7F}
{Federici}, A., {Arduino}, P., {Riva}, A., \& {Zerbi}, F.~M. 2012, in
  Astronomical Society of India Conference Series, Vol.~7, 7


\bibitem[{ORSA(2017)}]{github}
ORSA. 2017, {GitHub repository}, \url{https://github.com/orsa-unige}, [Online;
  accessed 26-November-2019]


\bibitem[{{Ricci} {et~al.}(2017){Ricci}, {Sada}, {Navarro-Meza},
  {L{\'o}pez-Valdivia}, {Michel}, {Fox Machado}, {Ram{\'o}n-Fox},
  {Ayala-Loera}, {Brown Sevilla}, {Reyes-Ruiz}, {La Camera}, {Righi}, {Cabona},
  {Tosi}, {Truant}, {Peterson}, {Prieto-Arranz}, {Velasco}, {Pall{\'e}}, \&
  {Deeg}}]{2017PASP..129f4401R}
{Ricci}, D., {Sada}, P.~V., {Navarro-Meza}, S., {L{\'o}pez-Valdivia}, R.,
  {Michel}, R., {Fox Machado}, L., {Ram{\'o}n-Fox}, F.~G., {Ayala-Loera}, C.,
  {Brown Sevilla}, S., {Reyes-Ruiz}, M., {La Camera}, A., {Righi}, C.,
  {Cabona}, L., {Tosi}, S., {Truant}, N., {Peterson}, S.~W., {Prieto-Arranz},
  J., {Velasco}, S., {Pall{\'e}}, E., \& {Deeg}, H. 2017, \pasp, 129, 064401


\bibitem[{{Righi}(2016)}]{2016NCimC..39..284R}
{Righi}, C. 2016, Nuovo Cimento C Geophysics Space Physics C, 39, 284


\bibitem[{{Sprimont} {et~al.}(2014){Sprimont}, {Ricci}, \&
  {Nicastro}}]{2014RMxAC..45...75S}
{Sprimont}, P.~G., {Ricci}, D., \& {Nicastro}, L. 2014, in Revista Mexicana de
  Astronomia y Astrofisica Conference Series, Vol.~45, 75


\end{thebibliography}

\end{document}